\documentclass[twoside,twocolumn,9pt]{article}
\usepackage{extsizes}

\usepackage[usedoi,linkdoi,super=false,usetitle=true,super]{rsc}
\usepackage[version=4]{mhchem}
\usepackage[center]{subfigure}
\usepackage[left=1.5cm, right=1.5cm, top=1.785cm, bottom=2.0cm]{geometry}
\usepackage{balance}
\usepackage{times,mathptmx}
\usepackage{sectsty}
\usepackage{graphicx}
\usepackage{lastpage}
\usepackage[format=plain,justification=justified,singlelinecheck=false,font={stretch=1.125,small,sf},labelfont=bf,labelsep=space]{caption}
\usepackage{float}
\usepackage{fancyhdr}
\usepackage{fnpos}
\usepackage[english]{babel}
\usepackage{comment}
\usepackage{tabularx}
\usepackage{amssymb,amsmath}
\usepackage{multirow}
\usepackage{booktabs}
\usepackage{longtable}
\usepackage{color}
\usepackage{comment}
\usepackage[colorlinks=true, allcolors=blue]{hyperref}
\usepackage[capitalize]{cleveref}
\usepackage{threeparttable}

\addto{\captionsenglish}{
  
}
\usepackage{array}
\usepackage{droidsans}
\usepackage{charter}
\usepackage[T1]{fontenc}
\usepackage[usenames,dvipsnames,table]{xcolor}
\usepackage{setspace}
\usepackage[compact]{titlesec}
\usepackage{hyperref}
\usepackage{bm}
\usepackage{mathtools}

\usepackage{listings} 
\usepackage{amsmath}
\usepackage{mathtools}
\usepackage{nicefrac,xspace}
\usepackage{tikz}

\usepackage{calrsfs}
\DeclareMathAlphabet{\pazocal}{OMS}{zplm}{m}{n}

\newcolumntype{C}{>{\centering\arraybackslash}X}
\newcolumntype{R}{>{\raggedleft\arraybackslash}X}
\newcolumntype{L}{>{\raggedright\arraybackslash}X}

\newcommand{\eg}{\emph{e.\,g.}}

\newcommand*\etal{\emph{et al.}}

\newcommand*\citeit[1]{\citeauthor{#1}\cite{#1}}

\newcommand*\new[1]{{\color{black}#1}}

\usepackage{epstopdf}

\definecolor{cream}{RGB}{222,217,201}

\DeclareMathAlphabet{\mathpzc}{OT1}{pzc}{m}{it}
\begin{document}

\pagestyle{fancy}
\thispagestyle{plain}
\fancypagestyle{plain}{

\fancyhead[C]{\includegraphics[width=18.5cm]{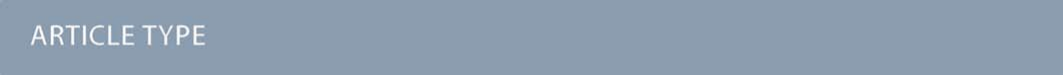}}
\fancyhead[L]{\hspace{0cm}\vspace{1.5cm}\includegraphics[height=30pt]{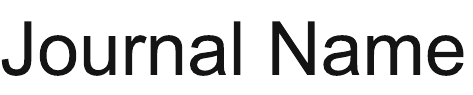}}
\fancyhead[R]{\hspace{0cm}\vspace{1.7cm}\includegraphics[height=55pt]{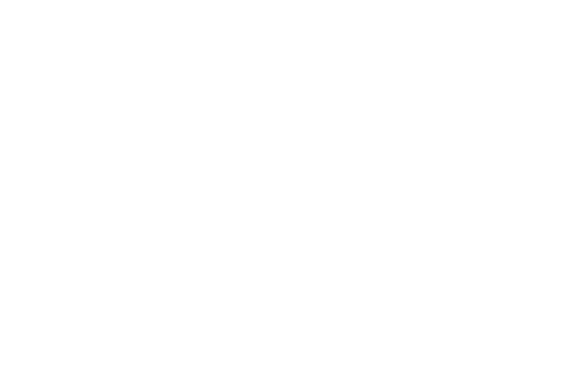}}
\renewcommand{\headrulewidth}{0pt}
}

\makeFNbottom
\makeatletter
\renewcommand\LARGE{\@setfontsize\LARGE{15pt}{17}}
\renewcommand\Large{\@setfontsize\Large{12pt}{14}}
\renewcommand\large{\@setfontsize\large{10pt}{12}}
\renewcommand\footnotesize{\@setfontsize\footnotesize{7pt}{10}}
\makeatother

\renewcommand{\thefootnote}{\fnsymbol{footnote}}
\renewcommand\footnoterule{\vspace*{1pt}%
\color{cream}\hrule width 3.5in height 0.4pt \color{black}\vspace*{5pt}}
\setcounter{secnumdepth}{5}

\makeatletter
\renewcommand\@biblabel[1]{#1}
\renewcommand\@makefntext[1]%
{\noindent\makebox[0pt][r]{\@thefnmark\,}#1}
\makeatother
\renewcommand{\figurename}{\small{Fig.}~}
\sectionfont{\sffamily\Large}
\subsectionfont{\normalsize}
\subsubsectionfont{\bf}
\setstretch{1.125} 
\setlength{\skip\footins}{0.8cm}
\setlength{\footnotesep}{0.25cm}
\setlength{\jot}{10pt}
\titlespacing*{\section}{0pt}{4pt}{4pt}
\titlespacing*{\subsection}{0pt}{15pt}{1pt}

\fancyfoot{}
\fancyfoot[LO,RE]{\vspace{-7.1pt}\includegraphics[height=9pt]{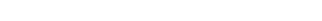}}
\fancyfoot[CO]{\vspace{-7.1pt}\hspace{13.2cm}\includegraphics{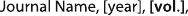}}
\fancyfoot[CE]{\vspace{-7.2pt}\hspace{-14.2cm}\includegraphics{head_foot/RF}}
\fancyfoot[RO]{\footnotesize{\sffamily{1--\pageref{LastPage} ~\textbar  \hspace{2pt}\thepage}}}
\fancyfoot[LE]{\footnotesize{\sffamily{\thepage~\textbar\hspace{3.45cm} 1--\pageref{LastPage}}}}
\fancyhead{}
\renewcommand{\headrulewidth}{0pt}
\renewcommand{\footrulewidth}{0pt}
\setlength{\arrayrulewidth}{1pt}
\setlength{\columnsep}{6.5mm}
\setlength\bibsep{1pt}

\makeatletter
\newlength{\figrulesep}
\setlength{\figrulesep}{0.5\textfloatsep}

\newcommand{\topfigrule}{\vspace*{-1pt}%
\noindent{\color{cream}\rule[-\figrulesep]{\columnwidth}{1.5pt}} }

\newcommand{\botfigrule}{\vspace*{-2pt}%
\noindent{\color{cream}\rule[\figrulesep]{\columnwidth}{1.5pt}} }

\newcommand{\dblfigrule}{\vspace*{-1pt}%
\noindent{\color{cream}\rule[-\figrulesep]{\textwidth}{1.5pt}} }

\makeatother

\twocolumn[
  \begin{@twocolumnfalse}
\vspace{3cm}
\sffamily
\begin{tabular}{m{4.5cm} p{13.5cm} }

\includegraphics{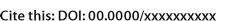} & \noindent\LARGE{\textbf{Density functional benchmark for quadruple hydrogen bonds$^\dag$}} \\
\vspace{0.3cm} & \vspace{0.3cm} \\

 & \noindent\large{Usman Ahmed,\textit{$^{a}$}
Mikael P.\ Johansson,\textit{$^{a,b}$}
Susi Lehtola,\textit{$^{a}$}
and Dage Sundholm$^{\ast,}$\textit{$^{a}$}}
\\

\includegraphics{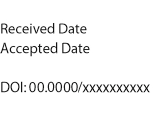} & \noindent\normalsize{%
    } \\

\end{tabular}

 \end{@twocolumnfalse} \vspace{0.6cm}

  ]

\renewcommand*\rmdefault{bch}\normalfont\upshape
\rmfamily
\section*{}
\vspace{-1cm}


\footnotetext{\textit{$^{a}$~Department of Chemistry, 
P.O. Box 55, A. I. Virtasen aukio 1, FI-00014 University of Helsinki, Finland.
E-mail: susi.lehtola@helsinki.fi, sundholm@chem.helsinki.fi}}
\footnotetext{\textit{$^{b}$~CSC-IT Center for Science Ltd., P.O. Box 405, FI-02101 Espoo, Finland; E-mail: mikael.johansson@csc.fi}}


\footnotetext{\dag~Electronic Supplementary Information (ESI) available: the
Libxc identifiers of the employed functionals, violin plots for \new{the density
functional approximations} ranked between 37 and 152, as well as reproduced
data from \citeit{D4CP02361G}: the employed TPSSh-D3/def2-TZVPP optimized
Cartesian coordinates of the studied molecules along with visualizations of the
molecular structures, as well as the coupled-cluster reference energies. See
DOI: 00.0000/00000000.}



\begin{abstract}
Hydrogen bonding is an important non-covalent interaction that plays a major
role in molecular self-organization and supramolecular structures.  It can be
described accurately with \textit{ab initio} quantum chemical wave function
methods, which become computationally expensive for large molecular assemblies.
Density functional theory (DFT) offers a better balance between accuracy and
computational cost, and can be routinely applied to large systems.  A large
number of density functional approximations (DFAs) has been developed, but
their accuracy depend on the application, necessitating benchmark studies to
guide their selection for use in applications. Some of us have recently
determined highly accurate hydrogen bonding energies of 14 quadruply
hydrogen-bonded dimers by extrapolating coupled-cluster energies to the
complete basis set limit as well as extrapolating electron correlation
contributions with a continued-fraction approach [U. Ahmed \etal, Phys. Chem.
Chem. Phys., 2024, 26, 24470].  In this work, we study the reproduction of
these bonding energies at the DFT level using 152 DFAs. The top ten
\new{density functional approximations} are composed of eight variants of the
Berkeley functionals both with and without dispersion corrections, and two
Minnesota 2011 functionals augmented with a further dispersion correction.  We
find the B97M-V functional with the non-local correlation functional replaced
by an empirical D3BJ dispersion correction to be the best \new{DFA}, while
changes to the dispersion part in other Berkeley functionals lead to poorer
performance in our study.
\end{abstract}

\section{Introduction}
\label{sec:intro}

Molecular self-assembly is a spontaneous process where disordered
molecules organize into ordered structures without external
influence.\cite{doi:10.1126/science.1962191,
  doi:10.1021/acs.chemrev.1c00884} The process leads to large
supramolecular architectures, which are complexed through non-covalent
interactions that guide the self-assembly. Understanding the complex
self-assembly mechanism is necessary for the design of large
supramolecular systems.

Hydrogen bonding is one of the most important non-covalent
interactions governing the self-assembly due to its strength,
selectivity, and directionality.\cite{B103906G,BUCKINGHAM20081}  The
significance of hydrogen bonding for stabilizing molecules and
biomolecules is well known from, \eg{}, its role in the structure of
deoxyribonucleic acid (DNA).  The strength of hydrogen bonding depends
strongly on the number of bonds, each hydrogen bond
increasing the strength of the bonding.  Molecules with four
hydrogen bonds per molecule are major examples of multiple
hydrogen bonding in self-organized systems. Self-complementarity
enables identical molecules to interact and form larger and uniform
supramolecular structures. Arrays of four hydrogen bonds can bind together in either a DDAA-AADD or a DADA-ADAD bonding motif, where D and A denote hydrogen bond donors and
acceptors, respectively. These motifs are particularly consequential for
designing stable assemblies with a given structure.

The quadruply hydrogen bonded systems having self-complementary
DDAA-AADD or DADA-ADAD motifs have been of scientific interest during
the past few decades.\cite{doi:10.1021/ja974112a,
  doi:10.1021/jo048223l, doi:10.1021/ja992830m, doi:10.1021/ja020098c,
  https://doi.org/10.1111/j.1432-1033.2004.04062.x,
  doi:10.1021/ja037312x, doi:10.1021/jo0508705,
  https://doi.org/10.1002/ejoc.201201450, doi:10.1021/jo301643z,
  C1NJ20162J, C0NJ00197J, B610239E,
  https://doi.org/10.1002/chem.201605208}
According to the traditional view, dimers with the DDAA-AADD motif have four
attractive secondary A$\cdots$D interactions leading to stronger hydrogen
bonding than in dimers with the DADA-ADAD motif, where all secondary
interactions have been considered to be repulsive.\cite{C3RA43814G} However, some of us
showed recently that dimers with the DDAA-AADD motif have five
attractive secondary interactions (A$\cdots$D and A$\cdots$A) and those with
the DADA-ADAD motifs have three attractive secondary interactions, because the
secondary acceptor-acceptor (A$\cdots$A) interactions are also
attractive.\cite{D4CP02361G,D4CP02889A}

Accurate computational analysis of the strength and stability of supramolecular
systems with hydrogen bonding is challenging, because the molecules are large.
While \textit{ab initio} correlated methods do offer an accurate description of
hydrogen bonding, the rapidly increasing computational costs hamper
computational studies of large molecules. Kohn--Sham density functional theory
(DFT)\cite{physrev.140.a1133} has become the method of
choice for studying such systems.\cite{doi:10.1063/1.4869598}  DFT is widely
used due to its cost effectiveness and availability in all modern quantum chemical
software packages. Since the exact exchange-correlation functional is not
known, more than 600 density functional approximations (DFAs) have so far been
proposed and implemented.\cite{Lehtola2018_S_1} Since the accuracy of the DFAs
varies depending on the application, it is not obvious which functional should
be used for studying molecules with multiple hydrogen bonds.

The recent work by \citeit{MonterodeHijes2024_JCP_131102} on the
density isobars of water and the melting temperature of ice based on
machine learned density functional potential energy surfaces
underlines the need for reinvestigating DFT models of hydrogen
bonds. Their study found small changes in the semi-empirical
dispersion correction in DFT calculations to result in large changes in the bulk
properties of water and ice. As multiple hydrogen bonding carries a
significant role for the structure of water, identifying functionals (which include dispersion corrections)
that work especially well for modeling highly hydrogen bonded systems
is useful to guide future studies of complex liquids and other systems
where hydrogen-bonding interactions play a major role.

The accuracy of various DFAs for modeling hydrogen bonds has been
evaluated in a number of works.\cite{A904357H, rabuck2000performance,
 Staroverov2003_JCP_12129, doi:10.1021/jp035869t, Zhao2005_JCTC_415} 
More recent
DFA benchmark studies of hydrogen-bonding strengths have employed the
S22, S66 and S66x8 datasets.\cite{doi:10.1021/ct6002719, Zhao2008_TCA_215,
  B600027D, doi:10.1021/ar700111a,
  https://doi.org/10.1002/cphc.201100826,Goerigk2017_PCCP_32184, 
  doi:10.1063/1.3545971, doi:10.1021/ct4003527, C3CP51559A,
  doi:10.1021/acs.jctc.7b00220, https://doi.org/10.1002/jcc.23263,
  https://doi.org/10.1002/jcc.20570, C6CP00688D, D2CP03938A,
  doi:10.1021/acs.jctc.9b01265, doi:10.1021/acs.jctc.0c00715}
The pronounced relevance of hydrogen bonding in chemistry and their DFT modeling is underlined by the fact that hydrogen bond strengths have been included in the fitting of the Minnesota
functionals since their first versions,\cite{Zhao2005_JCTC_415, Zhao:05b, Zhao:06} for example.

In this work, we study how well various DFAs reproduce the hydrogen bonding in
14 quadruply hydrogen-bonded dimers \new{with
\ce{N}$\cdots$\ce{H}$\cdots$\ce{O} interactions.} Half of the molecules exhibit
the DDAA-AADD \new{hydrogen-bonding motif} and the other half \new{has
the DADA-ADAD motif}.  We consider 152 DFAs including the Hartree--Fock (HF) method. The
employed dispersion correction schemes include Grimme's empirical dispersion
corrections (DFT-D3\cite{Grimme2011_JCC_1456},
DFT-D4\cite{Caldeweyher2019_JCP_154122}) and a non-local (NL) correlation
functional (VV10).\cite{Vydrov2010_JCP_244103}

Some of us have recently reported accurate reference values for the
molecules studied in this work, which were calculated with
domain-based local pair natural orbital coupled-cluster theory at the
singles, doubles and perturbative triples DLPNO-CCSD(T) level
extrapolated to the complete basis set limit.\cite{D4CP02361G}
\citeit{D4CP02361G} also extrapolated the reference energies towards
the Schr\"odinger limit following \citeit{Goodson2002_JCP_6948},
producing the data that we will use as reference in this work.

This work is outlined as follows.
The computational details are given in \cref{sec:methods} along with
the list of the studied DFAs.  The obtained results are presented in
\cref{sec:results}: basis-set convergence is discussed in
\cref{sec:convergence-results}, and the results are analyzed in
\cref{sec:benchmark-results}. The work is summarized and its main conclusions are drawn in
\cref{sec:conclusion}.

\section{Computational methods \label{sec:methods}}

\begin{figure}[H]
\centering
\subfigure[DADA01]{\includegraphics[width=0.25\columnwidth]{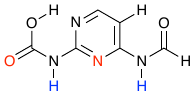}}
\subfigure[DADA02]{\includegraphics[width=0.25\columnwidth]{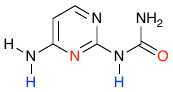}}
\subfigure[DADA03]{\includegraphics[width=0.25\columnwidth]{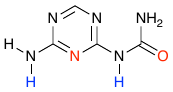}}
\\
\subfigure[DADA04]{\includegraphics[width=0.25\columnwidth]{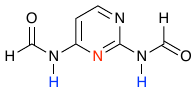}}
\subfigure[DADA05]{\includegraphics[width=0.25\columnwidth]{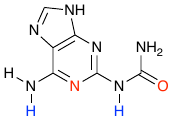}}
\\
\subfigure[DADA06]{\includegraphics[width=0.41\columnwidth]{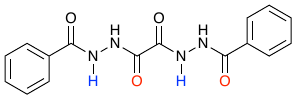}}
\subfigure[DADA08]{\includegraphics[width=0.25\columnwidth]{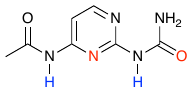}}
\caption{
\new{The monomers of the quadruple hydrogen-bonded dimers with 
DADA-ADAD hydrogen-bonding pattern. 
The previously studied DADA07\cite{D4CP02361G} is
omitted because the CCSD(T)-cf reference data are not available.  
The figures are made with Jmol\cite{jmol}.}} 
\label{fig:DADA_fig}
\end{figure}

\begin{figure}[H]
\centering
\subfigure[DDAA02]{\includegraphics[width=0.26\columnwidth]{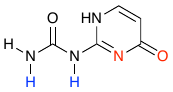} }
\subfigure[DDAA03]{\includegraphics[width=0.26\columnwidth]{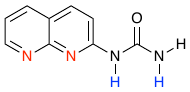} }
\subfigure[DDAA04]{\includegraphics[width=0.26\columnwidth]{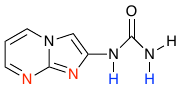} }
\\
\subfigure[DDAA05]{\includegraphics[width=0.26\columnwidth]{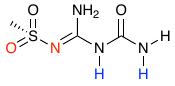} }
\subfigure[DDAA06]{\includegraphics[width=0.26\columnwidth]{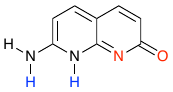} }
\\
\subfigure[DDAA07]{\includegraphics[width=0.26\columnwidth]{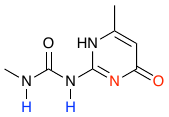} }
\subfigure[DDAA08]{\includegraphics[width=0.26\columnwidth]{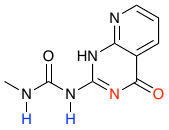} }
\caption{
\new{The monomers of the quadruple hydrogen-bonded dimers with the 
DDAA-AADD hydrogen-bonding pattern. 
The previously studied DDAA01\cite{D4CP02361G} is
omitted because the CCSD(T)-cf reference data are not available.  
The figures are made with Jmol\cite{jmol}.}} 
\label{fig:DDAA_fig}
\end{figure}

The benchmark considers the \new{14 dimers}  of the study of
\citeit{D4CP02361G} that determined hydrogen-bonding energies for
TPSSh-D3/def2-TZVPP optimized geometries at the DLPNO-CCSD(T) level of
theory extrapolated to the complete basis set limit, as well as
towards the Schr\"odinger limit using the continued-fraction (cf)
scheme of \citeit{Goodson2002_JCP_6948}. We denote the reference level
of theory as CCSD(T)-cf. The reference energies \new{and molecular structures 
of the dimers reported in the ESI\dag\ are taken from
\citeit{D4CP02361G}. The molecular 
structures of the monomers with the DADA and DDAA motifs are shown in
\cref{fig:DADA_fig,fig:DDAA_fig}, respectively. }  


The hydrogen-bonding energies were calculated in this work at the DFT
level of theory with versions 1.7 and 1.9 of the Psi4
program,\cite{Smith2020_JCP_184108} likewise employing the fixed
TPSSh-D3/def2-TZVPP geometries of \citeit{D4CP02361G}. We used the
systematic Karlsruhe family of basis sets to study the convergence of
the hydrogen-bonding energies towards the complete basis set limit. We
used split-valence polarized (def2-SVP), triple-$\zeta$ valence
polarized (def2-TZVP) or double valence polarized (def2-TZVPP), as
well as quadruple$-\zeta$ valence with a double set of polarization
functions (def2-QZVPP) basis sets in this study.\cite{Weigend:05} We
note here that since the considered molecules only contain the H, C,
N, O, and S atoms, the def2-QZVP basis is identical to def2-QZVPP and
only the latter is therefore considered, while the difference in def2-TZVP and
def2-TZVPP is limited to a larger basis set for H in def2-TZVPP. The effect of diffuse basis
functions was also studied with the corresponding basis sets
(def2-SVPD, def2-TZVPD, def2-TZVPPD, and
def2-QZVPPD).\cite{Rappoport:10}

\new{The hydrogen-bonding energies were corrected for basis set superposition
errors (BSSE) calculated using the counterpoise correction
method\cite{Jansen:69,Boys:70}.  The BSSE was obtained by comparing the
energies of the two monomers of the dimer structure calculated using the dimer
basis set to the monomer energies calculated using the monomer basis sets and
the same molecular structure.  The BSSE was added to the hydrogen-bonding
energy calculated for fully optimized monomers using the basis set of the
monomers.}

The calculations were performed using
the default (75,302) exchange-correlation grid in Psi4, which uses
75 radial grid points placed following \citeit{Treutler:95}, and an
angular Lebedev grid with 302 points.\cite{Lebedev:95} The VV10
non-local correlation functional\cite{Vydrov2010_JCP_244103} was
likewise evaluated on the corresponding (50,146) default grid. The
resolution-of-the-identity approximation was used in all calculations
with the universal Hartree--Fock auxiliary basis
sets.\cite{Weigend2008_JCC_167}

The 75 base DFAs considered in this work, which have been listed in
\cref{tab:functionals}, comprise 14 semi-local generalized gradient
approximation (GGA) functionals, 21 global hybrid (GH) GGA
functionals, 13 range-separated (RS) GGA functionals, 13 semi-local
meta-GGA (mGGA) functionals, 8 GH mGGA functionals, and 5 RS mGGA
functionals, as well as HF (a GH functional without a semi-local
exchange-correlation part). All of the density functionals were
evaluated with Libxc,\cite{Lehtola2018_S_1} as usual in
Psi4. Versions 6.0.0 and 6.2.2 of the library were employed. We note
that some functionals such as B3LYP, HSE03, and HSE06 have various definitions
in the literature,\cite{Lehtola2023_JCP_114116} which is avoided
through the use of the standard implementations in Libxc.
\Cref{tab:functionals} also shows the empirical
(DFT-D3\cite{Grimme2011_JCC_1456},
DFT-D4\cite{Caldeweyher2019_JCP_154122}) dispersion corrections or
non-local correlation (VV10\cite{Vydrov2010_JCP_244103}) additionally
considered with some of the functionals. Various combinations of these
dispersion models (or the lack thereof) with the 75 base DFAs yield
the 152 DFAs investigated in this study.

\new{The study by \citeit{D4CP02361G} showed that solvent effects increase the
hydrogen-bonding energy of the tautomers DADA08 and DDAA06 by only 4-5\%. We
have not considered any solvent effects in this work because they are
relatively small. Solvent effects are not expected to significantly affect the
ranking of the DFAs. }

\begin{table*}
  \caption{The density functional approximations considered in this
work. The three dispersion correction schemes are also indicated.
}
  \begin{threeparttable}
    \begin{tabular}{lccccc|lccccc}
      \hline
    functional & type & D3 & D4 & NL & ref. & functional & type & D3 & D4 & NL & ref. \tabularnewline
      \hline
B97-D3 & GGA & \citenum{Grimme2006_JCC_1787}$^a$ &  &  & \citenum{Grimme2006_JCC_1787} & 
$\omega$B97 & RS GGA & & & & \citenum{Chai2008_JCP_84106}                         \tabularnewline
BLYP & GGA & \citenum{Grimme2011_JCC_1456}$^a$ & \citenum{Caldeweyher2019_JCP_154122} & \citenum{Hujo2011_JCTC_3866} & \citenum{Becke1988_PRA_3098,Lee1988_PRB_785,Miehlich1989_CPL_200}                   &  
$\omega$B97X & RS GGA & & \citenum{Caldeweyher2019_JCP_154122} &  & \citenum{Chai2008_JCP_84106}     \tabularnewline
BOP & GGA & \citenum{Grimme2011_JCC_1456}$^a$ &  &  & \citenum{Tsuneda1999_JCP_10664}                &   
$\omega$B97X-D$^b$  & RS GGA  & & & & \citenum{Chai2008_PCCP_6615}                                                 \tabularnewline
BP86 &  GGA & \citenum{Goerigk2017_PCCP_32184}$^a$ &  & \citenum{Kesharwani2016_JCTC_444} & \citenum{Becke1988_PRA_3098, Perdew1986_PRB_8822}    &  
$\omega$B97X-V  & RS GGA & \citenum{Najibi2018_JCTC_5725}$^c$ & & &  \citenum{Mardirossian2014_PCCP_9904}             \tabularnewline
CHACHIYO &  GGA & & & & \citenum{Chachiyo2020_M_3485, Chachiyo2020_CTC_112669}                           &  
$\omega$PBE & RS GGA & \citenum{Grimme2011_JCC_1456}$^a$ &  &  & \citenum{Henderson2008_JCP_194105,Weintraub2009_JCTC_754}        \tabularnewline
KT1 &  GGA & & & & \citenum{Keal2003_JCP_3015}                                                         &  
CAM-B3LYP & RS GGA & \citenum{Grimme2011_JCC_1456}$^a$ &  &  & \citenum{Yanai2004_CPL_51}                                               \tabularnewline
KT2 &  GGA & & & & \citenum{Keal2003_JCP_3015}                                                         &   
CAMh-B3LYP & RS GGA & & & & \citenum{Shao2020_JCTC_587}                                           \tabularnewline
KT3 &  GGA & & &  &  \citenum{Keal2004_JCP_5654}                                                       &   
CAM-QTP-00 & RS GGA & & & & \citenum{Verma2014_JCP_18A534}                             \tabularnewline
N12 &  GGA & \citenum{Goerigk2015_JPCL_3891}$^a$ &  &  &  \citenum{Peverati2012_JCTC_2310}                                                &  
CAM-QTP-01 & RS GGA & & & & \citenum{Jin2016_JCP_34107}                                         \tabularnewline
PBE & GGA & \citenum{Grimme2011_JCC_1456}$^a$ & \citenum{Caldeweyher2019_JCP_154122} & \citenum{Kesharwani2016_JCTC_444} & \citenum{Perdew1996_PRL_3865, Perdew1997_PRL_1396}  & 
CAM-QTP-02 & RS GGA & & & & \citenum{Haiduke2018_JCP_184106}                                        \tabularnewline
revPBE & GGA & \citenum{Grimme2011_JCC_1456}$^a$ & \citenum{Caldeweyher2019_JCP_154122} & \citenum{Hujo2011_JCTC_3866} & \citenum{Zhang1998_PRL_890} & 
HSE03&RS GGA&\citenum{Grimme2011_JCC_1456}$^a$&\citenum{Caldeweyher2019_JCP_154122}&&\citenum{Heyd2003_JCP_8207,Heyd2006_JCP_219906}\tabularnewline
rPBE & GGA & \citenum{Grimme2011_JCC_1456}$^a$ & \citenum{Caldeweyher2019_JCP_154122} &  & \citenum{Hammer1999_PRB_7413}   & 
HSE06 & RS GGA & \citenum{Grimme2011_JCC_1456}$^a$ & \citenum{Caldeweyher2019_JCP_154122} &  & \citenum{Heyd2003_JCP_8207,Heyd2006_JCP_219906,Krukau2006_JCP_224106}  
\tabularnewline
PW91 & GGA & \citenum{Reimers2015_PNAS_6101}$^a$ & \citenum{Caldeweyher2019_JCP_154122} &  &  \citenum{Perdew1992_PRB_6671, Perdew1993_PRB_4978}   & 
N12-SX & RS GGA & \citenum{Goerigk2015_JPCL_3891}$^a$ &  &  &  \citenum{Peverati2012_PCCP_16187}       \tabularnewline
XLYP & GGA & \citenum{Grimme2011_JCC_1456}$^a$ & \citenum{Caldeweyher2019_JCP_154122} & & \citenum{Xu2004_PNASUSA_2673} & 
B97M-V & mGGA & \citenum{Najibi2018_JCTC_5725}$^c$ &  &  & \citenum{Mardirossian2015_JCP_74111}                                          \tabularnewline
HF & GH & \citenum{Grimme2011_JCC_1456}$^a$ & \citenum{Caldeweyher2019_JCP_154122} & & &
M06-L & mGGA & & & & \citenum{Zhao2006_JCP_194101}                                     \tabularnewline
B1LYP & GH GGA & \citenum{Grimme2011_JCC_1456}$^a$ & \citenum{Caldeweyher2019_JCP_154122} &  & \citenum{Adamo1997_CPL_242}   & 
revM06-L & mGGA & & & & \citenum{Wang2017_PNASUSA_8487}                                \tabularnewline
B3LYP & GH GGA & \citenum{Grimme2011_JCC_1456}$^a$& \citenum{Caldeweyher2019_JCP_154122}& \citenum{Hujo2011_JCTC_3866}& \citenum{Stephens1994_JPC_11623}& 
M11-L & mGGA & \citenum{Goerigk2015_JPCL_3891}$^a$ &  &  & \citenum{Peverati2012_JPCL_117}                                \tabularnewline
revB3LYP & GH GGA & & & & \citenum{Lu2013_CTC_64} & 
MN12-L & mGGA & \citenum{Goerigk2015_JPCL_3891}$^a$ &  &  & \citenum{Peverati2012_PCCP_13171}  \tabularnewline
B3PW91 & GH GGA & \citenum{Grimme2011_JCC_1456}$^a$ &  & \citenum{Hujo2011_JCTC_3866} & \citenum{Becke1993_JCP_5648} & 
MN15-L & mGGA & & & & \citenum{Yu2016_JCTC_1280}                                 \tabularnewline
B3P86 & GH GGA & \citenum{Goerigk2017_PCCP_32184}$^a$ &  &  & \citenum{Becke1993_JCP_5648}                        & 
MVS & mGGA & & & & \citenum{Sun2015_PNASUSA_685, Perdew2009_PRL_26403}       \tabularnewline
B97 & GH GGA & \citenum{Grimme2011_JCC_1456}$^c$ & & & \citenum{Becke1997_JCP_8554}           & 
SCAN & mGGA & \citenum{Grimme2011_JCC_1456}$^a$ & \citenum{Caldeweyher2019_JCP_154122} &  & \citenum{Sun2015_PRL_36402}  \tabularnewline
B97-1 & GH GGA & \citenum{Goerigk2017_PCCP_32184}$^a$ &  &  & \citenum{Hamprecht1998_JCP_6264}    & 
rSCAN & mGGA & & & & \citenum{Bartok2019_JCP_161101}                                 \tabularnewline
B97-2 & GH GGA & \citenum{Goerigk2017_PCCP_32184}$^a$ &  &  &  \citenum{Wilson2001_JCP_9233} & 
r\textsuperscript{2}SCAN & mGGA & \citenum{Grimme2011_JCC_1456}$^a$ & \citenum{Ehlert2021_JCP_61101} &  & \citenum{Furness2020_JPCL_8208,Furness2020_JPCL_9248}               \tabularnewline
B97-3 & GH GGA & & & &  \citenum{Keal2005_JCP_121103}                                                    & 
TASK & mGGA & & & & \citenum{Aschebrock2019_PRR_33082, Perdew1992_PRB_13244}               \tabularnewline
BHandH & GH GGA & & & & \citenum{Becke1993_JCP_1372}                                             &  
TPSS & mGGA & \citenum{Grimme2011_JCC_1456}$^a$ & \citenum{Caldeweyher2019_JCP_154122} & \citenum{Hujo2011_JCTC_3866} & \citenum{Tao2003_PRL_146401, Perdew2004_JCP_6898}   \tabularnewline
BHandHLYP &  GH GGA & & & &  \citenum{Becke1988_PRA_3098, Becke1993_JCP_1372}   & 
revTPSS & mGGA & \citenum{Grimme2011_JCC_1456}$^a$ & \citenum{Caldeweyher2019_JCP_154122} & \citenum{Kruse2018_JCTC_95} & \citenum{Perdew2009_PRL_26403, Perdew2011_PRL_179902}     \tabularnewline
BMK & GH GGA & \citenum{Grimme2011_JCC_1456}$^a$ &  &  & \citenum{Boese2004_JCP_3405}   & 
M06 & GH mGGA & & &  & \citenum{Zhao2008_TCA_215} \tabularnewline
HCTH120 & GH GGA & \citenum{Grimme2011_JCC_1456}$^a$ &  &  &  \citenum{Boese2000_JCP_1670}       & 
revM06 & GH mGGA & & & & \citenum{Wang2018_PNASUSA_10257}                                 \tabularnewline
HCTH407 & GH GGA & \citenum{Grimme2011_JCC_1456}$^a$ &  &  & \citenum{Boese2001_JCP_5497}                           & 
M06-2X & GH mGGA & & & & \citenum{Zhao2008_TCA_215}                                                         \tabularnewline
MPW1PW& GH GGA & \citenum{Grimme2011_JCC_1456}$^a$ & \citenum{Caldeweyher2019_JCP_154122} &  & \citenum{Adamo1998_JCP_664} & 
M08-HX & GH mGGA & & & & \citenum{Zhao2008_JCTC_1849}   \tabularnewline
O3LYP & GH GGA & \citenum{Grimme2011_JCC_1456}$^a$ & \citenum{Caldeweyher2019_JCP_154122} &  & \citenum{Hoe2001_CPL_319,Cohen2001_MP_607} & 
M08-SO & GH mGGA & & & & \citenum{Zhao2008_JCTC_1849}                                               \tabularnewline
PBE0 & GH GGA & \citenum{Grimme2011_JCC_1456}$^a$ & \citenum{Caldeweyher2019_JCP_154122} & \citenum{Kesharwani2016_JCTC_444} &  \citenum{Adamo1999_JCP_6158, Ernzerhof1999_JCP_5029} & 
MN15 & GH mGGA & \citenum{Grimme2011_JCC_1456}$^a$ &  &  & \citenum{Yu2016_CS_5032}                      \tabularnewline
revPBE0 & GH GGA & \citenum{Grimme2011_JCC_1456}$^a$ & \citenum{Caldeweyher2019_JCP_154122} & \citenum{Hujo2011_JCTC_3866} &  \citenum{Zhang1998_PRL_890,Perdew1996_PRL_3865,Perdew1997_PRL_1396} & 
TPSSh & GH mGGA & \citenum{Grimme2011_JCC_1456}$^a$ & \citenum{Caldeweyher2019_JCP_154122} &  & \citenum{Staroverov2003_JCP_12129}  \tabularnewline
QTP17 & GH GGA & & & &  \citenum{Jin2018_JCP_64111}                  & 
revTPSSh & GH mGGA & \citenum{Grimme2011_JCC_1456}$^a$ & \citenum{Caldeweyher2019_JCP_154122} &  & \citenum{Csonka2010_JCTC_3688} \tabularnewline
SOGGA11-X & GH GGA & \citenum{Goerigk2015_JPCL_3891}$^a$ &  &  & \citenum{Peverati2011_JCP_191102}   & 
$\omega$B97M-V & RS mGGA & \citenum{Najibi2018_JCTC_5725}$^c$ &  &  & \citenum{Mardirossian2016_JCP_214110}                                                    \tabularnewline
X3LYP & GH GGA & \citenum{Grimme2011_JCC_1456}$^a$ & \citenum{Caldeweyher2019_JCP_154122} &  & \citenum{Xu2004_PNASUSA_2673}       & 
M06-SX & RS mGGA & & & & \citenum{Wang2020_PNASUSA_2294}                                        \tabularnewline
&&&&&&M11 & RS mGGA & \citenum{Grimme2011_JCC_1456}$^a$ &  &  & \citenum{Peverati2011_JPCL_2810}  \tabularnewline
&&&&&&revM11 & RS mGGA & & & & \citenum{Verma2019_JPCA_2966}                                                  \tabularnewline
&&&&&&MN12-SX & RS mGGA & \citenum{Goerigk2015_JPCL_3891}$^a$ &  &  & \citenum{Peverati2012_PCCP_16187}                                      \tabularnewline
   \hline
\end{tabular}
   \begin{tablenotes}
   \item [$^a$] Empirical dispersion with Becke--Johnson\cite{Johnson2005_JCP_24101} 3-body damping.
   \item [$^b$] The functional was parametrized with Grimme's original dispersion correction, which is not available in Psi4, so the calculations with this functional do not include dispersion.
   \item [$^c$] Empirical dispersion with Becke--Johnson\cite{Johnson2005_JCP_24101} 2-body damping.
   \end{tablenotes}
  \end{threeparttable}
  \label{tab:functionals}
\end{table*}

\section{Results \label{sec:results}}

\subsection{Basis-set convergence}
\label{sec:convergence-results}

\begin{figure}
\centering
  \subfigure[Convergence without diffuse functions]{
    \includegraphics[width=1.00\columnwidth]{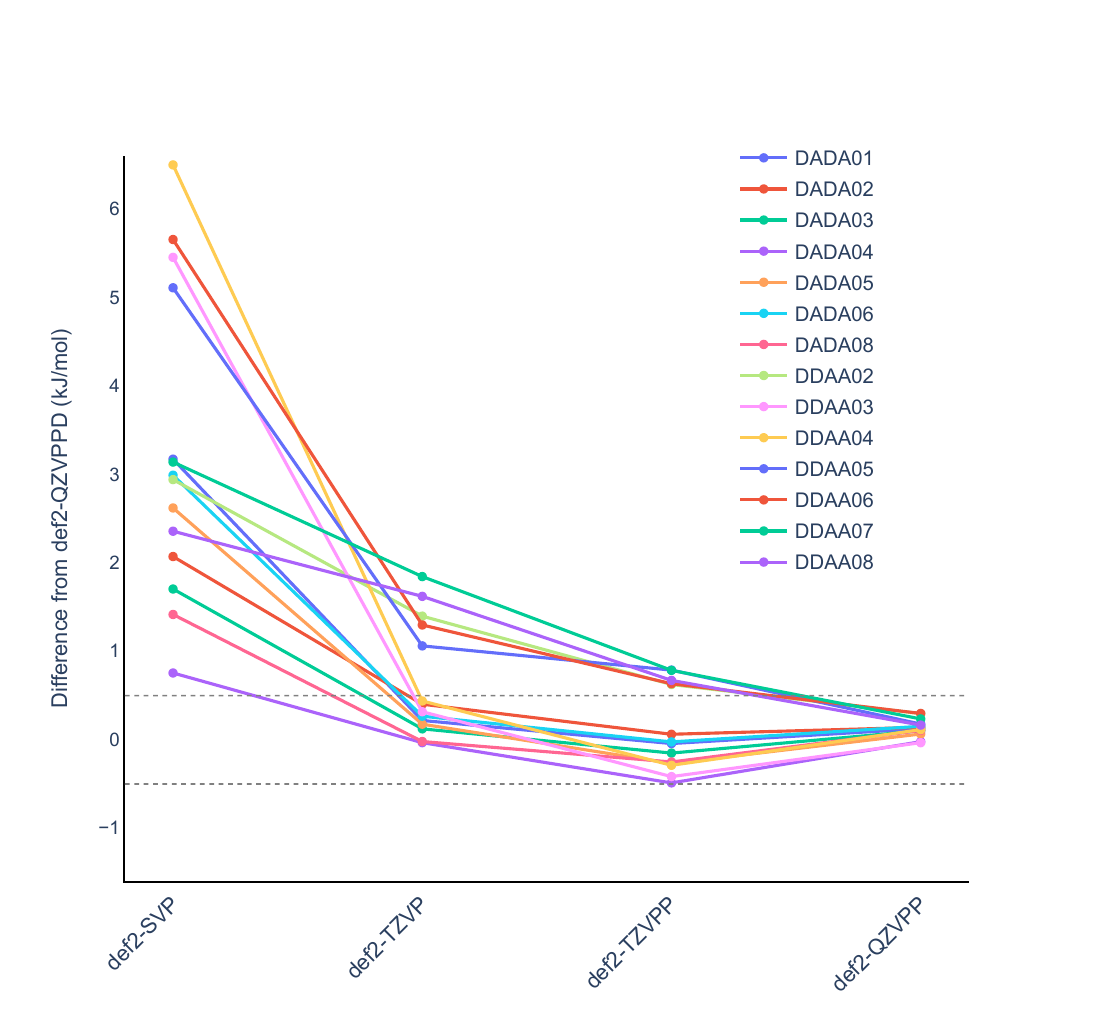}
    \label{fig:basis-convergence-a}}

  \subfigure[Convergence with diffuse functions]{
    \includegraphics[width=1.0\columnwidth]{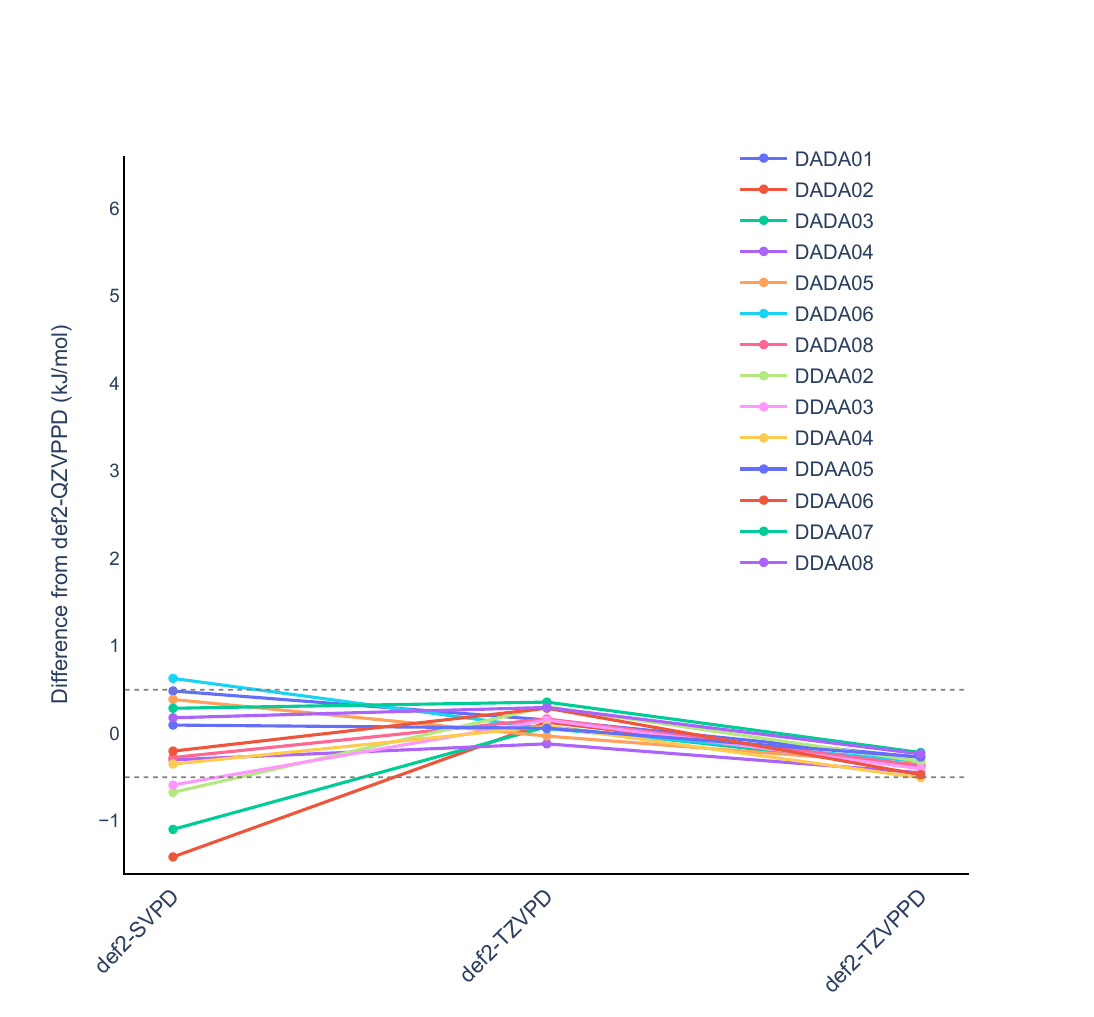}
    \label{fig:basis-convergence-b}}

  \caption{Basis-set convergence plot of the hydrogen-bonding
    energies, relative to B97M-D3/def2-QZVPPD values. The figures show
    an energy window of $\pm 0.5$ kJ/mol as horizontal dashed lines.
\label{fig:basis-convergence}
}
\end{figure}

We start the analysis by determining the basis-set convergence of the
hydrogen bonding energies. We use the B97M-D3 functional and use the
largest basis set considered (def2-QZVPPD) to determine reference
values for this part of the study. The resulting basis-set truncation
errors in the hydrogen-bonding energies are shown in
\cref{fig:basis-convergence} as a function of basis set.

\Cref{fig:basis-convergence-a} shows the convergence for the basis
sets without diffuse functions. The truncation errors decrease
systematically, as the basis set is increased from split-valence with
polarization to polarized valence triple- and quadruple-$\zeta$.
The convergence is relatively slow with the cardinal
number, and not all triple-$\zeta$ results are not yet converged to
$\pm 0.5$ kJ/mol precision.

\Cref{fig:basis-convergence-b} shows the analogous results including
diffuse functions. The range of truncation errors for the
polarized split-valence basis sets drops significantly from $\approx$ 0.8--7
kJ/mol with def2-SVP in \cref{fig:basis-convergence-a} to $\approx$
$-1.5$--0.5 kJ/mol with def2-SVPD; this highlights the need to include
diffuse functions for describing non-covalent bonding, as in the
present hydrogen bonded systems. A convergence to within $\pm 0.5$
kJ/mol precision with respect to the basis set is already achieved
with the def2-TZVPD basis set. It is thus clear from the data in
\cref{fig:basis-convergence} that the def2-QZVPPD basis set should
afford benchmark accuracy hydrogen-bond energies suitable for the
present study, and we will use the def2-QZVPPD basis set for the rest
of the study.

\subsection{The best \new{density functional approximations} 
\label{sec:benchmark-results}}

The \new{density functional approximations} were ranked based on their mean
absolute errors (MAE).  The ranking is shown in
\cref{tab:ranking-2,tab:ranking-3} together with the corresponding MAEs and
maximum absolute errors (MAX).  The lowest MAE is obtained with the B97M-D3
functional, which is the B97M-V functional where the VV10 nonlocal correlation
functional (indicated by -V in the functional's name) has been replaced with an
empirical D3 dispersion correction; the original B97M-V functional is ranked
4$^\mathrm{th}$. The second-lowest MAE is achieved with the $\omega$B97M-V
functional, which is a RS GGA functional again with VV10 dispersion. Replacing
VV10 with the D3 dispersion correction drops the ranking to the 7$^\mathrm{th}$
place.

\begin{table*}
\caption{The mean absolute errors (MAE) and maximum absolute errors (MAX) for
the hydrogen-bonding strength of the 14 studied molecules (in kJ/mol) using the
76 highest ranked \new{density functional approximations}. The reference
hydrogen-bonding energies are extrapolated CCSD(T)-cf values from the
literature.\cite{D4CP02361G} The \new{DFAs} are ordered in increasing MAE. }
  \label{tab:table2}
  \begin{threeparttable}
\begin{tabularx}{\textwidth}{LLRR|LLRR}
\hline
Rank & Functional & MAE & MAX & Rank & Functional & MAE & MAX \tabularnewline
\hline
1 & B97M-D3 & 1.82 & 5.24 & 39 & BLYP-D4 & 6.36 & 14.27 \tabularnewline
2 & $\omega$B97M-V & 2.06 & 5.45 & 40 & TPSS-D3 & 6.37 & 12.98 \tabularnewline
3 & M11-L-D3 & 2.13 & 5.22 & 41 & M08-HX & 6.43 & 10.81 \tabularnewline
4 & B97M-V & 2.17 & 6.71 & 42 & $\omega$B97X-D4 & 6.67 & 11.55 \tabularnewline
5 & $\omega$B97 & 2.21 & 5.86 & 43 & revTPSS-NL & 6.73 & 12.11 \tabularnewline
6 & $\omega$B97X-V & 2.43 & 6.09 & 44 & r$^2$SCAN-D4 & 7.07 & 13.55 \tabularnewline
7 & $\omega$B97M-D3 & 2.48 & 5.24 & 45 & TPSSh-D3 & 7.15 & 12.97 \tabularnewline
8 & $\omega$B97X-D3 & 3.09 & 6.25 & 46 & r$^2$SCAN-D3 & 7.29 & 13.63 \tabularnewline
9 & $\omega$B97X & 3.11 & 7.98 & 47 & HF-D4 & 7.31 & 14.14 \tabularnewline
10 & M11-D3 & 3.17 & 7.27 & 48 & PBE-D3 & 7.65 & 14.64 \tabularnewline
11 & revTPSSh-D3 & 3.20 & 7.55 & 49 & PW91-D3 & 7.70 & 14.23 \tabularnewline
12 & SCAN & 3.43 & 8.91 & 50 & revM11 & 7.98 & 13.51 \tabularnewline
13 & revTPSS-D3 & 3.48 & 9.21 & 51 & TPSS-D4 & 7.99 & 15.50 \tabularnewline
14 & revPBE-NL & 3.64 & 8.94 & 52 & B1LYP-D3 & 8.12 & 12.75 \tabularnewline
15 & revTPSS-D4 & 3.65 & 9.64 & 53 & M06-L & 8.26 & 13.41 \tabularnewline
16 & revTPSSh-D4 & 3.99 & 8.98 & 54 & N12-SX & 8.29 & 25.27 \tabularnewline
17 & XLYP-D3 & 4.21 & 10.57 & 55 & MN12-SX-D3 & 8.45 & 14.75 \tabularnewline
18 & $\omega$PBE-D3 & 4.24 & 12.62 & 56 & BMK-D3 & 8.49 & 14.50 \tabularnewline
19 & rPBE-D4 & 4.30 & 10.82 & 57 & CAM-QTP-01 & 8.54 & 20.80 \tabularnewline
20 & B97-D3 & 4.45 & 12.54 & 58 & BLYP-NL & 8.66 & 13.94 \tabularnewline
21 & BLYP-D3 & 4.58 & 11.96 & 59 & TPSSh-D4 & 8.82 & 14.81 \tabularnewline
22 & XLYP-D4 & 4.72 & 11.76 & 60 & PBE-D4 & 9.26 & 16.42 \tabularnewline
23 & BOP-D3 & 4.73 & 12.09 & 61 & HSE03 & 9.35 & 29.92 \tabularnewline
24 & HCTH407-D3 & 4.78 & 15.06 & 62 & SCAN-D4 & 9.75 & 16.54 \tabularnewline
25 & revPBE-D4 & 4.83 & 12.10 & 63 & M06 & 9.85 & 16.87 \tabularnewline
26 & HCTH120-D3 & 4.84 & 14.53 & 64 & B3PW91-D3 & 9.92 & 16.30 \tabularnewline
27 & rPBE-D3 & 5.03 & 16.16 & 65 & M11 & 9.98 & 14.92 \tabularnewline
28 & revPBE0-D3 & 5.05 & 8.89 & 66 & B3LYP-D3 & 10.05 & 14.89 \tabularnewline
29 & M06-2X & 5.12 & 9.73 & 67 & B1LYP-D4 & 10.09 & 14.61 \tabularnewline
30 & revPBE-D3 & 5.37 & 15.66 & 68 & MPW1PW-D3 & 10.14 & 16.95 \tabularnewline
31 & revPBE0-NL & 5.42 & 11.61 & 69 & MN12-L-D3 & 10.25 & 17.57 \tabularnewline
32 & CAM-QTP-00 & 5.53 & 15.23 & 70 & PW91-D4 & 10.48 & 17.06 \tabularnewline
33 & rSCAN & 5.56 & 16.54 & 71 & PBE-NL & 10.65 & 16.64 \tabularnewline
34 & B97-1-D3 & 5.59 & 10.42 & 72 & MPW1PW-D4 & 10.74 & 17.73 \tabularnewline
35 & MVS & 5.66 & 13.53 & 73 & HSE06 & 10.74 & 31.54 \tabularnewline
36 & B97-2-D3 & 5.89 & 10.47 & 74 & PBE0-D3 & 10.82 & 17.40 \tabularnewline
37 & r$^2$SCAN & 5.93 & 15.69 & 75 & SCAN-D3 & 10.88 & 17.64 \tabularnewline
38 & revPBE0-D4 & 6.01 & 11.03 & 76 & TPSS-NL & 10.88 & 16.88 \tabularnewline
\hline
\end{tabularx}
  \end{threeparttable}
  \label{tab:ranking-2} \\
\end{table*}

\begin{table*}
\caption{The mean absolute errors (MAE) and maximum absolute errors (MAX) for
the hydrogen-bonding strength of the 14 studied molecules (in kJ/mol) using the
\new{density functional approximations} ranked from 77 to 152. The reference
hydrogen-bonding energies are extrapolated CCSD(T)-cf values from the
literature.\cite{D4CP02361G} The \new{DFAs} are ordered in increasing MAE. }
  \label{tab:table3}
\begin{tabularx}{\textwidth}{LLRR|LLRR}
\hline
Rank & Functional & MAE & MAX & Rank & Functional & MAE & MAX \tabularnewline
\hline
77 & B3LYP-D4 & 11.05 & 15.90 & 115 & BMK & 19.84 & 31.71 \tabularnewline
78 & BP86-D3 & 11.10 & 19.82 & 116 & X3LYP & 19.96 & 40.69 \tabularnewline
79 & revM06 & 11.20 & 15.90 & 117 & N12 & 20.22 & 48.75 \tabularnewline
80 & X3LYP-D3 & 11.26 & 16.17 & 118 & CAMh-B3LYP & 20.44 & 41.43 \tabularnewline
81 & HSE06-D3 & 11.35 & 17.78 & 119 & $\omega$B97X-D & 20.59 & 42.63 \tabularnewline
82 & B3PW91-NL & 11.60 & 18.99 & 120 & MN12-L & 20.79 & 27.11 \tabularnewline
83 & X3LYP-D4 & 11.68 & 16.45 & 121 & MPW1PW & 21.09 & 47.74 \tabularnewline
84 & \mbox{CAM-B3LYP-D3} & 11.78 & 17.57 & 122 & TPSSh & 23.77 & 50.62 \tabularnewline
85 & PBE0-D4 & 11.90 & 18.31 & 123 & KT3 & 23.86 & 51.05 \tabularnewline
86 & CAM-B3LYP & 12.10 & 29.56 & 124 & $\omega$PBE & 23.91 & 48.60 \tabularnewline
87 & M08-SO & 12.13 & 19.50 & 125 & revTPSSh & 24.32 & 46.15 \tabularnewline
88 & \mbox{SOGGA11-X-D3} & 12.13 & 17.82 & 126 & TPSS & 25.91 & 53.91 \tabularnewline
89 & B3LYP-NL & 12.18 & 17.31 & 127 & B3LYP & 26.54 & 50.90 \tabularnewline
90 & PBE0-NL & 12.22 & 19.09 & 128 & revTPSS & 26.68 & 49.15 \tabularnewline
91 & HSE03-D3 & 12.53 & 19.42 & 129 & BP86 & 27.33 & 58.38 \tabularnewline
92 & MN15-D3 & 12.64 & 18.31 & 130 & B1LYP & 28.11 & 51.93 \tabularnewline
93 & PW91 & 12.80 & 35.00 & 131 & B97-0 & 28.81 & 53.34 \tabularnewline
94 & PBE0 & 13.04 & 34.52 & 132 & O3LYP-D4 & 29.21 & 51.57 \tabularnewline
95 & HSE06-D4 & 13.12 & 19.58 & 133 & O3LYP-D3 & 29.77 & 52.14 \tabularnewline
96 & MN15 & 13.33 & 19.31 & 134 & B3PW91 & 29.82 & 61.28 \tabularnewline
97 & MN15-L & 13.33 & 19.31 & 135 & QTP-17 & 30.03 & 47.11 \tabularnewline
98 & CAM-QTP-02 & 13.66 & 26.81 & 136 & B97-3 & 30.95 & 53.53 \tabularnewline
99 & M06-SX & 13.71 & 19.06 & 137 & KT1 & 31.33 & 39.56 \tabularnewline
100 & B3P86-D3 & 14.21 & 20.56 & 138 & B97-2 & 33.77 & 64.10 \tabularnewline
101 & HSE03-D4 & 14.26 & 20.90 & 139 & HCTH120 & 36.29 & 65.40 \tabularnewline
102 & BHANDHLYP & 15.21 & 33.09 & 140 & revPBE0 & 37.87 & 69.50 \tabularnewline
103 & SOGGA11-X & 15.59 & 31.93 & 141 & XLYP & 38.30 & 65.73 \tabularnewline
104 & M11-L & 15.69 & 26.51 & 142 & BLYP & 39.54 & 69.55 \tabularnewline
105 & BP86-NL & 16.09 & 23.11 & 143 & KT2 & 39.62 & 49.78 \tabularnewline
106 & N12-SX-D3 & 16.34 & 21.96 & 144 & BHANDH & 44.53 & 61.65 \tabularnewline
107 & HF-D3 & 16.63 & 25.87 & 145 & HF & 46.07 & 71.35 \tabularnewline
108 & PBE & 17.27 & 40.79 & 146 & HCTH407 & 47.72 & 79.14 \tabularnewline
109 & revM06-L & 17.67 & 26.51 & 147 & revPBE & 50.92 & 88.16 \tabularnewline
110 & B3P86 & 17.84 & 43.49 & 148 & rPBE & 51.03 & 85.81 \tabularnewline
111 & B97-1 & 17.96 & 36.89 & 149 & O3LYP & 61.36 & 102.66 \tabularnewline
112 & N12-D3 & 18.19 & 25.07 & 150 & BOP & 63.69 & 105.40 \tabularnewline
113 & revB3LYP & 19.61 & 41.55 & 151 & CHACHIYO & 63.76 & 111.80 \tabularnewline
114 & MN12-SX & 19.79 & 25.07 & 152 & TASK & 148.78 & 216.80 \tabularnewline
\hline
\end{tabularx}
  \label{tab:ranking-3} \\
\end{table*}

The Minnesota M11-L-D3 holds 3$^\mathrm{rd}$ place on the list with the added
D3 dispersion corrections, while the original functional parametrized without
the empirical dispersion term (M11-L) holds the 104$^\mathrm{th}$ place.

The \new{DFAs} ranked $5^\mathrm{th}$ to $9^\mathrm{th}$ are all Berkeley
$\omega$B97 (RS GGA) functionals with different dispersion correction terms.
$\omega$B97X-D3 is the $\omega$B97X-V\cite{Mardirossian2014_PCCP_9904}
functional with a D3
Becke--Johnson\cite{Johnson2005_JCP_24101,Grimme2011_JCC_1456} dispersion
correction instead of VV10\cite{Najibi2018_JCTC_5725}, and it should not to be
confused with the $\omega$B97X-D3 functional of \citeit{Lin2013_JCTC_263} which
was not studied in this work.

The Minnesota M11 functional with an additional D3 correction is on the
10$^\mathrm{th}$ place, while the original functional without the dispersion
term ranks 65$^\mathrm{th}$.

The $\omega$B97X-D4 functional is ranked 42$^\mathrm{nd}$, which is 34
places below the rank of $\omega$B97X-D3. This is likely because
$\omega$B97X-D4 is based on a different base functional than
$\omega$B97X-D3: it is built on top of an older functional,
$\omega$B97X instead of
$\omega$B97X-V.\cite{Caldeweyher2019_JCP_154122}
While a dispersion correction was included in the fitting of the $\omega$B97X-V
functional through the VV10 functional, $\omega$B97X was fit without dispersion
terms.

Various mGGA functionals and one revised GGA functional are ranked
11$^\mathrm{th}$ to 16$^\mathrm{th}$.

Five of the six studied Berkeley functionals are ranked among the 10 best
performing DFAs.  mGGA and global hybrid (GH) GGA functionals also appear among
the 20 most accurate functionals for hydrogen-bonding energies.
The best \new{DFAs at the GGA level} are ranked
17$^\mathrm{th}$ (XLYP-D3),
19$^\mathrm{th}$ (rPBE-D4), 21$^\mathrm{st}$ (BLYP-D3) and
22$^\mathrm{nd}$ (XLYP-D4).

\begin{figure*}
  \subfigure[]{
    \includegraphics[width=1.00\textwidth]{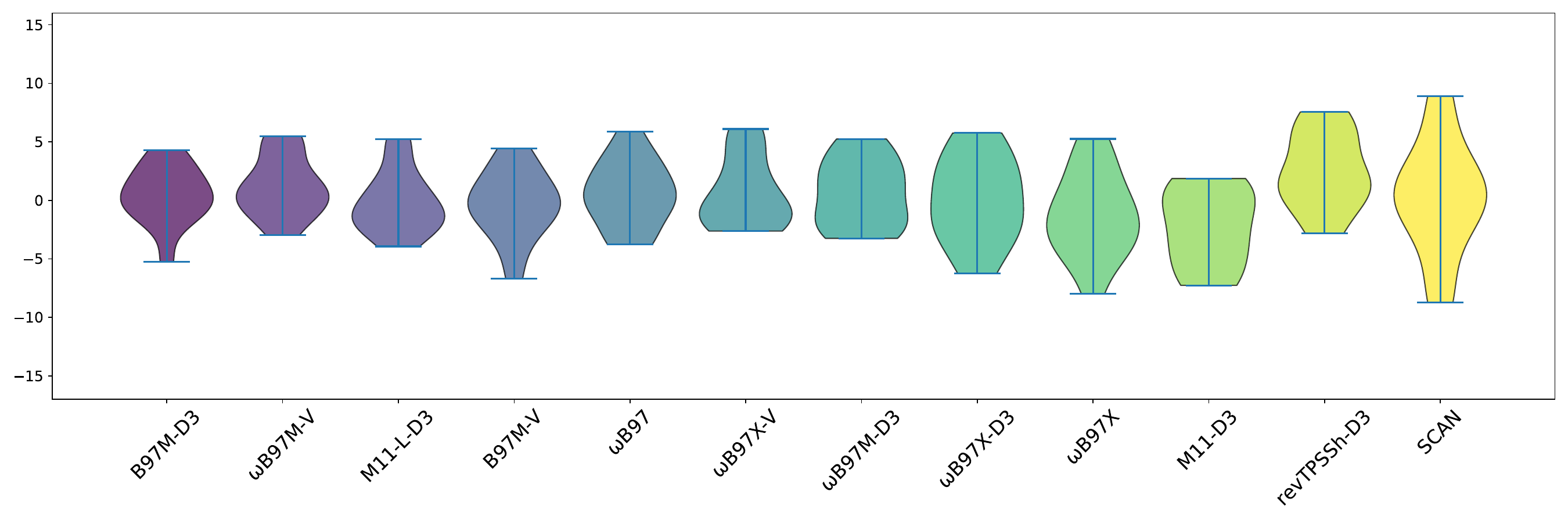}
    \label{fig:violins-a}}

  \subfigure[]{
    \includegraphics[width=1.0\textwidth]{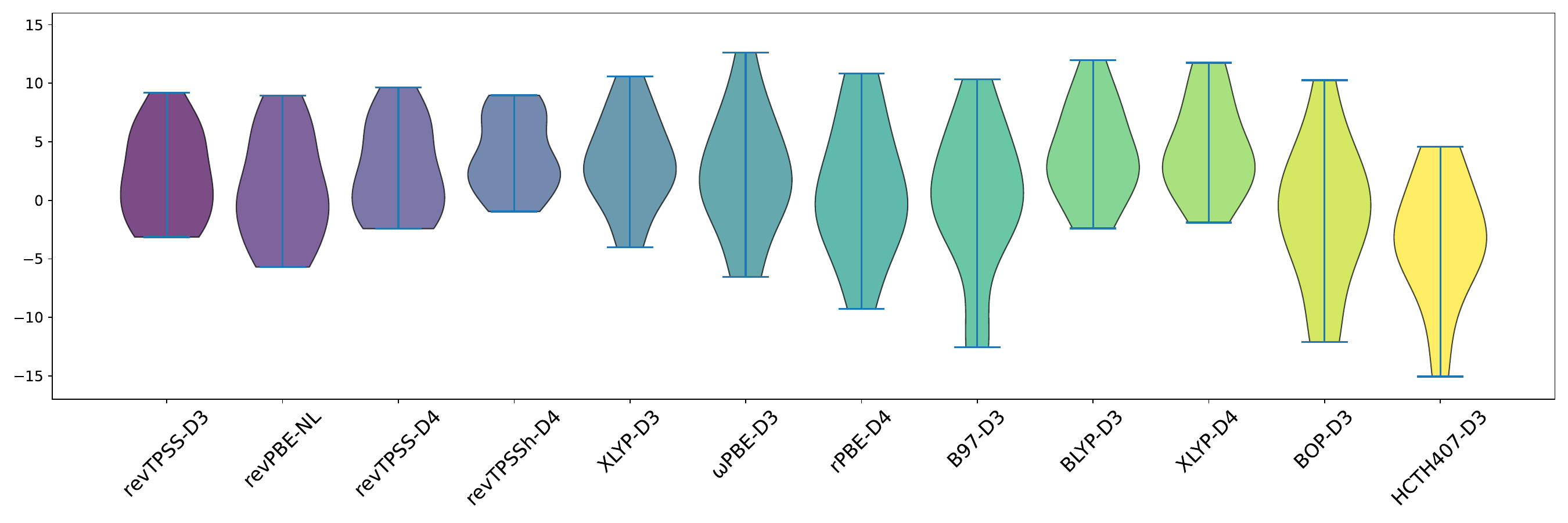}
    \label{fig:violins-b}}

  \subfigure[]{
    \includegraphics[width=1.0\textwidth]{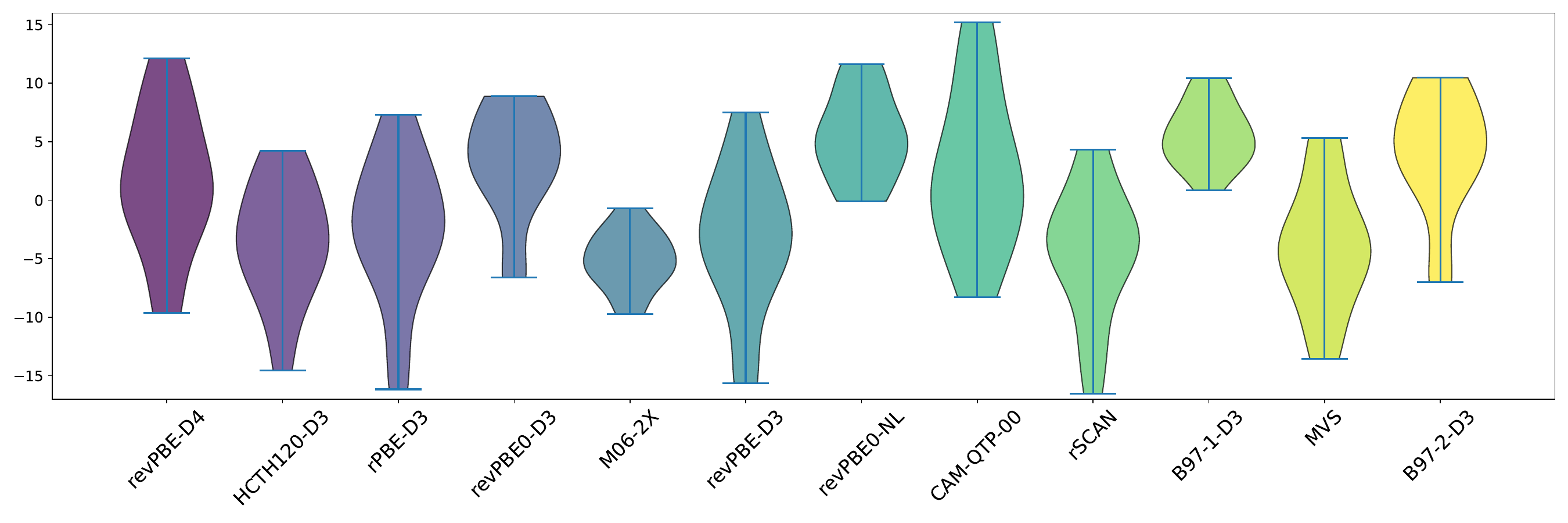}
    \label{fig:violins-c}}

\caption{Violin plots showing the error distributions in the
  hydrogen-bonding energies (in kJ/mol) for the investigated DFAs. The
  reference hydrogen-bonding energies are extrapolated CCSD(T)-cf values
  from the literature.\cite{D4CP02361G} }
  \label{fig:violins}
\end{figure*}

The error distributions for the hydrogen-bonding energies of the 36
top \new{DFAs} are shown as violin plots in \cref{fig:violins}, while
violin plots for the remaining \new{DFAs} are included in the
\new{ESI\dag}.  The average errors in the
hydrogen-bonding energies for the top \new{DFAs} are close to zero,
except for M11-D3, whose distribution is almost completely on the
negative side, implying that it systematically underestimates the
hydrogen bonding energies.  Similar plots are shown for the
\new{DFAs} ranked 13$^\mathrm{th}$--24$^\mathrm{th}$ in
\cref{fig:violins-b}, where the distributions are broader, and the
average is larger than zero for many \new{DFAs}, which implies that
the hydrogen-bonding energies are overestimated.  The error
distributions for the \new{DFAs} ranked between 25$^\mathrm{th}$ and
36$^\mathrm{th}$ shown in \cref{fig:violins-c} are broader and more
scattered than for the higher-ranked \new{DFAs}.  The violin plots
for the rest of the \new{DFAs} are given in the ESI\dag.

\new{Hydrogen-bonding energies have been previously calculated in the basis-set
limit at the second-order M{\o}ller-Plesset perturbation theory (MP2)
level\cite{D4CP02889A}. Since all deviations have the same sign, MP2
systematically overestimates the hydrogen-bonding energies, whereas the
deviations of the hydrogen-bonding energies calculated using the top DFAs are
both positive and negative. The hydrogen-bonding energies at the MP2 level
correlate linearly with the reference energies.  The angular coefficient of the
linear fit of the MP2 energies is $1.022\pm0.02$ with an offset of $-5.4\pm3.3$
kJ/mol. The MAE and MAX values at the MP2 level are \mbox{8.3 kJ/mol} and 14.6
kJ/mol, respectively, corresponding to the 55$^\mathrm{th}$ place in the
ranking list.  }


\subsection{Meta-GGA functionals}

The best performing \new{DFAs at the mGGA level} are B97M-D3 (1$^\mathrm{st}$), M11-L-D3
(3$^\mathrm{rd}$) and B97M-V (4$^\mathrm{th}$). SCAN and revTPSS-D3 are mGGA
functionals that also perform well, holding rank 12 and 13, respectively. SCAN
exhibits worse accuracy when paired with the D3 (ranked 75$^\mathrm{th}$) or D4
(ranked 62$^\mathrm{th}$) dispersion, suggesting overbinding. In contrast, the
revTPSS functional improves its performance significantly by adding the
empirical dispersion correction term to the functional, the rank increasing
from 125$^\mathrm{th}$ to 13$^\mathrm{th}$ (D3) or 15$^\mathrm{th}$ (D4), which
suggests that the hydrogen bonds are too weak at the revTPSS level, which is
rectified by the dispersion correction. The rest of the mGGA functionals among
the first third part of the ranking list includes rSCAN (33$^\mathrm{rd}$), MVS
(35$^\mathrm{th}$) and r$^2$SCAN (37$^\mathrm{th}$).  However, by considering
the dispersion interaction using the D3 correction term, the performance of
r$^2$SCAN drops from 37$^\mathrm{th}$ to 46$^\mathrm{th}$ and to
44$^\mathrm{th}$ with D4. Common mGGA functionals such as M06-L
(53$^\mathrm{rd}$) and MN15-L (97$^\mathrm{th}$) are in the second third part
of the ranking list, whereas M11-L without dispersion (104$^\mathrm{th}$),
revM06-L (109$^\mathrm{th}$), MN12-L (120$^\mathrm{th}$), TPSS
(126$^\mathrm{th}$), revTPSS (128$^\mathrm{th}$) and TASK (152$^\mathrm{nd}$)
are in last third of the ranking list.  The performance of the TPSS, revTPSS,
MN12-L, and MN12-L functionals improves significantly when adding the empirical
dispersion term.  The accuracy is in a few cases improved by using D4 instead
of D3, whereas using the VV10 dispersion correction leads to a lower accuracy
than obtained with the D3 correction.

\subsection{Range-separated hybrid functionals}

The Berkeley range-separated functionals perform well. The $\omega$B97M-V is
the best RS mGGA functional in the 2$^\mathrm{nd}$ place and the Minnesota M11
functional combined with the D3 correlation term is ranked 10$^\mathrm{th}$.
The best of the rest of the RS mGGA functionals is revM11, which is ranked
50$^\mathrm{th}$.  Three of the five RS GGA functionals from Berkeley are
ranked among the 10 best \new{DFAs}, while adding a dispersion
correction to $\omega$B97X leads to worse performace and 42$^\mathrm{nd}$ rank.
The $\omega$B97X-D functional performs poorly in our study
(119$^\mathrm{th}$), but this is likely caused by lack of dispersion
in our calculations with this functional; see \cref{tab:functionals}.


The best RS GGA functional from Gainesville (QTP) is
32$^\mathrm{nd}$, whereas the rest of the RS GGA functionals are ranked between
54$^\mathrm{th}$ and 135$^\mathrm{th}$. The short-range-only HSE03
and HSE06 functionals rank 61$^\mathrm{st}$ and 73$^\mathrm{rd}$, respectively. 
Adding the dispersion D3 term to
CAM-B3LYP has a very small effect on its MAE, whereas D3 improves its
performance by decreasing the MAX value by 40\%.

\subsection{Global hybrid functionals}

We have investigated the performance of 21 global hybrid GGA
functionals and added the D3 dispersion terms to 16 of them. Their
rank without the D3 term is in general 100$^\mathrm{th}$ and
larger. The only exceptions is PBE0 (94$^\mathrm{th}$). GH GGA
functionals perform better when adding the D3 dispersion term. The
best GH GGA functionals with the D3 correction are B97
(20$^\mathrm{th}$), HCTH407 (24$^\mathrm{th}$), HCTH120
(26$^\mathrm{th}$), revPBE0 (28$^\mathrm{th}$), B97-1
(34$^\mathrm{th}$) and B97-2 (36$^\mathrm{th}$). The ranking of the
other GH GGA-D3 functionals is 52$^\mathrm{nd}$ and larger. The use of
D4 or NL dispersion corrections leads in general to a slightly less
accurate functional than with D3.

The best global hybrid mGGA functional without additional dispersion term is
M06-2X in the 29$^\mathrm{th}$ place. The M08-HX functional is ranked
41$^\mathrm{st}$, whereas the rest of the GH mGGA functionals is ranked between
63$^\mathrm{rd}$ and 125$^\mathrm{th}$.  The performance of the TPSSh
functional improves from 122$^\mathrm{nd}$ to 45$^\mathrm{th}$ (D3) and
59$^\mathrm{th}$ (D4) when adding the empirical dispersion term to the
functional. Adding an empirical dispersion correction to the revTPSSh
functional improves its performance significantly: the ranking of revTPSSh
increases from 125$^\mathrm{th}$ to 11$^\mathrm{th}$ (revTPSSh-D3) or
16$^\mathrm{th}$ (revTPSSh-D4).

\new{MAX values generally increase with increasing MAE value. However,
revPBE0-D3 and M06-2X have much smaller MAX values than some of the higher
ranked DFAs.  Judged from the MAX value, revPBE0-D3 and M06-2X would be ranked
12$^\mathrm{th}$ and 17$^\mathrm{th}$ instead of 28$^\mathrm{th}$ and
29$^\mathrm{th}$ based on the MAE value, respectively.  }

\subsection{GGA functionals}

The revPBE-NL and XLYP-D3 functionals are the best performing \new{DFAs} among
the GGAs. Their performance drops from the 14$^\mathrm{th}$ place to
124$^\mathrm{th}$, and 17$^\mathrm{th}$ rank to 141$^\mathrm{th}$ without
dispersion corrections, respectively. The rPBE-D4 and BLYP-D3
functionals are the next ones on the GGA ranking list ranked 19$^\mathrm{th}$ and 21$^\mathrm{st}$, respectively.
The BOP functional places 23$^\mathrm{rd}$.
Other GGA functionals
among the first 50 on the ranking list are PBE-D3 and PW91-D3, which are ranked
48$^\mathrm{th}$ and 49$^\mathrm{th}$, respectively. The rest of the GGA
functionals including dispersion effects appear among the second third on the
ranking lists.  The ranking of HF is 145$^\mathrm{th}$, which improves to
46$^\mathrm{th}$ and 107$^\mathrm{th}$ with D3 and D4 dispersion corrections,
respectively.

\section{Summary and conclusions \label{sec:conclusion}}

We calculated the hydrogen-bonding strength for \new{14 dimers} using 152
density functional approximations (DFAs). The accuracy of the
calculated bond strengths with various DFAs was assessed by comparison
to the literature CCSD(T)-cf energies of \citeit{D4CP02361G}, which they
extrapolated to the complete basis set limit, as well as to the
Schr\"odinger limit using the continued-fraction methodology of
\citeit{Goodson2002_JCP_6948}.

We began the work by validating the numerical reliability of our computational
method by assessing the convergence of the hydrogen-bonding strengths across
the Karlsruhe basis set family for the B97M-D3 functional. Taking the
def2-QZVPPD values as reference, we observed that diffuse functions are
important to obtain converged interaction energies, and we therefore recommend
their use in modeling hydrogen bonds. As the def2-TZVPD and def2-TZVPPD values
are within 0.5 kJ/mol of the def2-QZVPPD data, we can safely assume that the
quadruple-$\zeta$ reference calculations are indeed of benchmark quality. The
def2-QZVPPD basis set is often used for benchmarking in the literature, and has
been employed in the fitting of the $\omega$B97M-V
functional,\cite{Mardirossian2016_JCP_214110} for example.

Having validated the numerical methodology, we continued with the benchmark. We
found the top of the list to be dominated by the Berkeley functionals, either
with or without dispersion.  The most accurate \new{DFA} for hydrogen-bonding
strengths was found to be B97M-D3, which is the B97M-V functional whose VV10
dispersion correction has been replaced by the empirical D3 dispersion
correction. We found the accuracy of mGGA functionals to be surprisingly
sensitive to the description of the dispersion energy, tentatively supporting
the recent findings of \citeit{MonterodeHijes2024_JCP_131102}.  Other
\new{DFAs} that accurately predicted hydrogen-bonding strengths included the
range-separated $\omega$B97M-V mGGA functional, the $\omega$B97, $\omega$B97X,
$\omega$B97X-V range-separated GGA functionals, as well as the semi-local
M11-L-D3 mGGA and the range-separated M11-D3 mGGA functional.

\subsection*{Acknowledgement}

This work has been supported by the Academy of Finland through project
numbers 314821, 340583, 350282, and 353749. It has also been
financially supported by the Magnus Ehrnrooth Foundation, the Oskar
{\"O}flund Foundation, and by the Swedish Cultural Foundation in
Finland. We acknowledge computational resources from CSC - IT Center
for Science, Finland.

\bibliographystyle{rsc}
\bibliography{literature,susi}
\newpage
\noindent
{\bf Graphical abstract}

\begin{figure}[H]
\centerline{\includegraphics[width=1.00\linewidth]{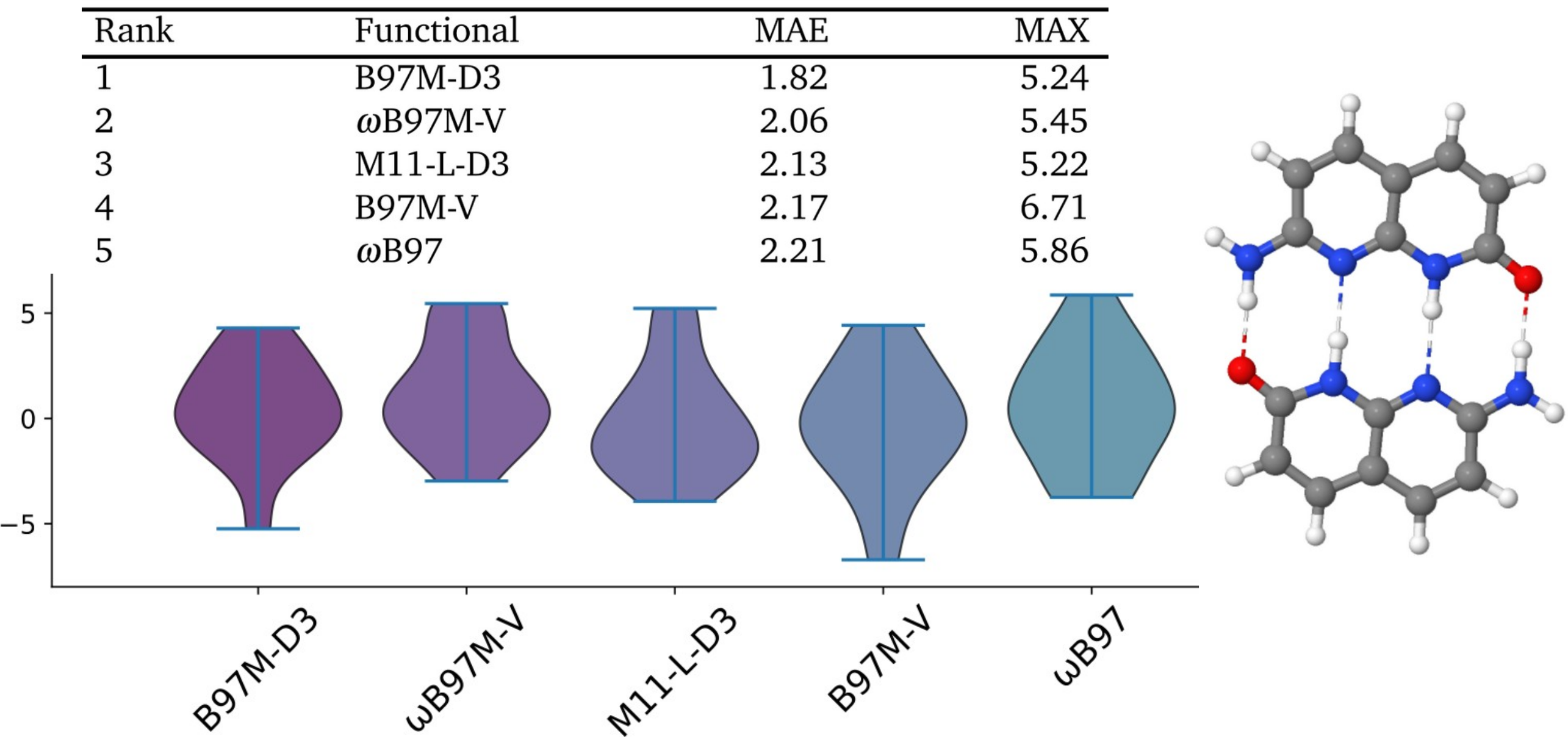}}

\noindent Hydrogen-bonding strengths of quadruply bonded
self-complementary dimers are calculated using 152 functionals and
compared to coupled-cluster reference energies extrapolated to the
Schrödinger limit.
\end{figure}

\end{document}